# Polarons and confinement of electronic motion to two dimensions in a layered transition metal oxide

H.M. Rønnow[*], Ch. Renner[†], G. Aeppli[†], T. Kimura[#] and Y. Tokura[§]

*Laboratory for Neutron Scattering, ETH-Zürich and Paul Scherrer Institut, 5232 Villigen, Switzerland*

[†]*London Centre for Nanotechnology & Department of Physics and Astronomy, University College London, Gower Street, London WC1E 6BT, UK*

[#]*Los Alamos National Laboratory, P.O. Box 1663, Los Alamos, NM 87545, USA*

[§]*Department of Applied Physics, University of Tokyo, Bunkyo-ku, Tokyo 113-8656, Japan.*

**A very remarkable feature of the layered transition metal oxides (TMO's), whose most famous members are the high-temperature superconductors (HTS's), is that even though they are prepared as bulk three-dimensional single crystals, they display hugely anisotropic electrical and optical properties, seeming to be insulating perpendicular to the layers and metallic within them. This is the phenomenon of confinement, a concept at odds with the conventional theory of solids and recognized[1] as due to magnetic and electron–lattice interactions in the layers which must be overcome at a substantial energy cost if electrons are to be transferred between layers. The associated energy gap or 'pseudogap' is particularly obvious in experiments where charge is moved perpendicular to the planes, most notably scanning tunneling microscopy (STM)[2] and polarized infrared spectroscopy.[3] Here, using the same experimental tools, we show that there is a second family of TMO's – the layered manganites $La_{2-2x}Sr_{1+2x}Mn_2O_7$ (LSMO) – with even more extreme confinement and pseudogap effects. The data, which are the first to resolve atoms in any metallic manganite, demonstrate quantitatively that because they are attached to polarons - lattice and spin textures within the planes -, it is equally difficult to remove carriers from the planes via vacuum tunneling into a conventional metallic tip, as it is for them to move between Mn-rich layers within the material itself.**

The manganese perovskite oxides (manganites) have been the subject of a decade-long revival, spurred by the colossal magnetoresistance[4,5] of thin films and sustained by the large variety of physical phenomena – including Jahn-Teller distortions, charge, spin and orbital ordering[6-9] – manifested in carefully prepared single crystals. A fruitful new area has been the layered manganites, the most prominent of which is LSMO. They are built from bilayers cut from the pseudocubic parent compound $(La,Sr)MnO_3$ and separated by rock-salt type lanthanide and alkaline-earth ion oxide sheets in a tetragonal crystal structure (Fig. 1a). These materials can be viewed as self-assembled stacks of two-dimensional electronic liquids weakly coupled to each other. They display the same type of anisotropic electric resistance (Fig. 1b) as the HTS, and tunneling magnetoresistance as in top-down fabricated devices, where switching from the low to high resistance states occurs via conversion from antiparallel to parallel configurations of the spins in adjacent bilayers.[10] The lamellar structure of some HTS cuprates has been very advantageous for experiments which are traditionally suitable for surface science, namely photoemission and STM. Much less information of this type is available for the manganites and especially their layered polymorphs. In this paper we report atomic scale microscopy and temperature-dependent spectroscopy of unprecedentedly large and (for manganites) flat surfaces of LSMO, providing definitive evidence for confinement as well as a rejection – for this two-dimensional metal – of the popular concept of electronic phase separation[11] where different electronic states coexist in adjacent regions.

As a function of Sr doping $x$, LSMO exhibits the familiar features of many manganites. Upon cooling from the high temperature paramagnetic state, LSMO develops a charge-ordered insulator around $x=0.5$ and a variety of ferro- and antiferromagnetic phases between $x=0.45$ and $x=0.30$ (Ref. 12). Here, we investigate the $x=0.30$ compound, whose outstanding feature, shared with optimally and underdoped HTS, is that in the paramagnetic phase, the electrical resistance within the planes appears metallic – decreasing with decreasing temperature – while that corresponding to hopping between them is much larger and insulating (Fig. 1b). Below $T_c = 90$ K, the material orders as an antiferromagnetic stack of two-dimensional ferromagnets (the bilayers), and the electrical properties in zero field track each other, following the paradigm of dirty metals with quantum corrections.[13]

We performed tunneling measurements with a STM in an ultra-high vacuum chamber (pressure $5 \cdot 10^{-10}$ Torr) and whose unique feature – in contrast to the instruments typically used in analogous studies of the cuprates – is that temperature could be controlled from 5 to 400K. The STM tips were chemically etched tungsten wires, and the LSMO single crystals were grown using the same floating zone technique and growth conditions[14] as for prior neutron measurements.[10] In-situ cleavage at room temperature produced clean LSMO surfaces which topographic tunneling images (Fig. 2a) revealed to be extremely flat, with a peak-to-peak roughness below 0.3Å over micron-sized terraces. Since 3D perovskites like $LaMnO_3$ do not cleave, LSMO is most likely to break between the double perovskite layers, in agreement with the half unit-cell high steps separating the terraces (Fig. 2a). Yet, despite atomically smooth surfaces, atomic resolution remains elusive. Whenever the atomic lattice was resolved, it was limited to tiny islands containing approximately 30 unit cells (Fig. 2b). These regions appear slightly raised above the background terrace, by too small an amount to be ascribed to adsorbed impurity atom clusters. They more likely correspond to regions where the tunneling conductance at 0.8V is slightly enhanced due to the local charge configuration and dynamics, and do not necessarily imply that the density of states is different in these regions. The observed lattice parameters are consistent with the ab-plane Mn ion spacing of 3.86Å. The atomic scale contrast in all of these tiny regions is very stable and reproducible over many hours (Fig. 2c-e). This low density of small and stable islands should not be regarded as phase separation, but as a local perturbation of the otherwise homogeneous system generated by some atomic scale defect.

We now turn to tunneling spectroscopy. For each temperature, we measured and analysed thousands of current-voltage characteristics, $I(V)$, at different locations over 10 to $10^6$ nm$^2$ areas. Spatially resolved tunneling spectroscopy gives answers to two questions about these materials: (1) Is there electronic phase separation into metallic and insulating regions[15,16] and (2) is there a specific signature associated with the ferromagnetic metal-insulator transition at $T_c$ = 90 K? Several conclusions can be directly read from the representative data depicted in Figs. 3a and 3b. First, the current distribution for each bias (grey-scale) stemming from 2000 spectra exactly matches the spread in each individual spectrum (dots). This means that the spectra are extremely homogeneous over micron-sized regions at all temperatures. Furthermore, it can be seen that spectra taken at 283 K from the tiny regions where atoms are resolved have the same spectroscopic signature as the surrounding featureless areas. Our most remarkable discovery is also directly visible in Fig. 3: the spectrum is gapped at 44 K, in striking contrast to what is suggested by the bulk transport measurements[17] showing a two orders of magnitude drop in resistivity to metallic behaviour below 90 K.

To quantify the tunneling characteristics, we first evaluated the zero-bias conductance, $\sigma_0$, namely the slopes at $V=0$ of thousands of spectra for each temperature found from fitting the low bias portion of the $I(V)$ curves to an ohmic background (red line in Figs. 3a and 3b). The slopes fall on a normal distribution, which gradually shifts away from zero and broadens upon increasing the temperature from 44 to 298 K (Fig. 3c, inset). The increasing spread as a function of temperature is accounted for by a constant electronic noise enhanced by the temperature dependence of the STM's piezoelectric coefficients, meaning that the broadening is a purely instrumental effect, and not a sign of increasing disorder with increasing temperature. Our data unequivocally demonstrate the absence of electronic phase separation into metallic and insulating regions at any temperature. Indeed, such electronic phase separation would manifest itself as a bimodal splitting of the normal distributions, which is not observed.

The temperature dependence of $\sigma_0$ is consistent with a thermally activated conductance across a gap $\Delta$ at the Fermi energy (Fig. 3c). Fitting $\sigma_0(T)$ to $T^{3/2} exp(-\Delta/2k_BT)$, we obtain $\Delta$ =196 ± 12 meV, in agreement with the gap $\Delta_\rho$=188 meV found from fitting the high temperature c-axis resistivity to a thermally activated form $\rho_c \propto T^{-3/2} exp(\Delta_\rho/2k_BT)$ (see Fig. 1b). Because the detailed nature of the tunnel junction may change between experiments, the absolute value of the zero-bias slope may vary (Fig. 3c). However, it does not affect the amplitude of the gap extracted from the fit, giving additional confidence that we are probing a gap intrinsic to LSMO.

While $\sigma_0$ can be determined in a robust manner, it is more difficult to derive the precise value of $\Delta$ from the tunneling spectra. In the absence of a model for the local density of states, we devised a method to extract two phenomenological parameters $V_-$ and $V_+$ related to $\Delta$ (Fig. 3d). Bearing in mind that the zero bias slope is simply due to thermal activation, $V_-$ ($V_+$) is defined as the negative (positive) sample bias at which the tunneling current departs by a given amount (threshold) from the ohmic background. The gap amplitude defined as $\Delta = V_+ - V_-$ depends on the particular choice of threshold, but its dramatic temperature dependence, doubling between room and base temperatures, much less so. Surprisingly, a more conventional behaviour is found for the chemical potential, $\mu = \frac{1}{2}(V_+ + V_-)$, which shows a linear dependence on temperature $\mu = \beta T$. Assuming the classic law $\mu = \frac{3}{4} k_B T \ln(m_h/m_e)$ for an intrinsic semiconductor,[18] the slope $\beta$ equals to $-\frac{3}{4}k_B ln(m_h/m_e)$, with $ln(m_h/m_e) = -1$ implying that holes travel more easily along the c axis than electrons ($m_h$ ($m_e$) is the hole (electron) effective mass).

Optical reflectivity measurements are another method for establishing the electronic excitation spectra of solids, and have the advantage over such probes as photoemission and scanning tunneling spectroscopy (STS) that they are much less surface-sensitive, while benefiting from insensitivity to the details – crucial for dc conductivity data - of electrical current flow in a macroscopic sample with electrical contacts. Consistency between optical data and STS is therefore a very definitive demonstration of whether the STS measurements represent surface or intrinsic bulk phenomena. Accordingly, Fig. 4 shows the optical conductivity[19] for LSMO samples. When polarized photons drive the carriers to move within the planes, the low-energy part of the spectra displays a dramatic Drude-like tail below $T_c$, signalling metallic behaviour (Fig. 4a), in excellent agreement with the in-plane resistivity data of Fig. 1b. Even at high energy, the in-plane optical conductivity shows significant temperature dependence with the broad peak around 1 eV shifting to lower energy with decreasing temperature. While the in-plane optical conductivity becomes increasingly metallic on cooling, the c-axis optical conductivity (Fig. 4b) shows a clear gap with no spectral weight up to about 400-500 meV (except for sharp phonon bands in the range 20-80 meV). The onset energy, corresponding to a filled to empty state transition in the density of states for the LSMO, is entirely consistent with the low temperature gap $\Delta = V_+ - V_-$ deduced from our STS spectra, shown in Fig. 3a. In complete contrast to the in-plane optical conductivity, there is absolutely no signature of a metallic Drude-like peak at low energy.

What is especially striking about our findings is the consistency of the STS and c-axis optical conductivity. Furthermore, the same gap value characterizes the temperature-dependent spectroscopy and the high temperature $\rho_c$, indicating that both measurements probe the same bulk property. However, on cooling we uncover the curious result that the STS results seem oblivious to the Néel point (90 K), below which $\rho_c$ displays seemingly metallic behaviour while the zero-bias conductance, the STS spectra and $\sigma_c(\omega)$ are all consistent with an insulator. How can we reconcile the striking discrepancy below $T_c$ between macroscopic resistivity measurements and the vacuum tunneling and optical spectroscopies where electrons must move perpendicular to the MnO planes? The most obvious idea is to assert, as has been done based on analysis of X-ray data for samples exposed to air, that the surface probed via tunneling is somehow insulating and non-magnetic,[20] and therefore different from the bulk sampled in electrical measurements. However, this is at odds with spin-polarised scanning electron microscopy clearly demonstrating that the top layer is ferromagnetic,[21] the in-plane metallic screening (see ref.22 and our more detailed discussion below) underlying the inability to resolve more than a few atoms on the smooth surfaces prepared by in-situ cleaves in our ultra-high vacuum STM experiments, static Secondary Ion Mass Spectroscopy (SIMS) showing that the top 1 nm of our cleaved sample has the same chemical composition as the bulk (see supplementary material for details), and by the general agreement of STS and optical conductivity, which is much less surface sensitive than tunneling.

The data show that whatever dominates the dc conductivity along c has no effect on the optical and STS measurements. To account for the dc metallicity, we consider LSMO as a stack of alternating metallic and insulating layers. In a perfect sample, the c-axis transport would always be thermally activated, with $\rho_c$ diverging at low temperature in agreement with the STM experiments. The sudden drop of $\rho_c$ at the ferromagnetic transition must then be due to a parallel conduction mechanism. The idea is that ferromagnetic domains with the spin orientation either pointing upwards or downwards form in each bilayer (Fig. 1a). Hence, while adjacent bilayers have magnetizations which are predominantly antiparallel, there will be defects where the domain walls in adjacent layers do not exactly coincide, resulting in regions containing adjacent bilayers whose magnetizations are parallel. Independent of spin orientation, tunneling is the main channel for c-axis conductivity, but the tunneling probability in the latter regions will be enhanced compared to the surrounding AF correlated regions, as also demonstrated by the large and negative c-axis magnetoresistance seen in the low temperature limit for this material.[10,12] While the defective regions will only contribute weakly to $\rho_c$ above $T_c$, before three-dimensional antiferromagnetism sets in, they turn into a well-connected network below $T_c$, effectively shunting the cation layers. Indeed, as depicted in Fig. 1b, a simple model with an activated intrinsic c-axis resistance $\rho_c^0(T) \propto T^{-3/2} \exp(\Delta_\rho/2k_BT)$, determined from the high-temperature portion of the measured curve, in parallel with an in-plane resistance, which is the product of $\rho_b(T)$ and a prefactor $\alpha(T)$, reproduces the temperature dependent conductivity $\rho_c(T)$. The prefactor is determined by the distances that the carriers travel within the planes, as well as the probability of moving carriers between bilayers in the defective regions where adjacent bilayers have parallel magnetizations. Because the latter quantity grows with the magnetic order parameter $M$ appearing below $T_c$, we use the Ansatz that $\alpha(T) = aM^2(T) + b$. The upshot is that as $\rho_c^0(T)$ diverges on lowering the temperature, the intrinsic anisotropy becomes larger and the c-axis transport increasingly mimics the in-plane resistivity. In contrast, the parallel channel is inoperative in the STM tunneling process, where a vacuum barrier is maintained between the tip and the sample surface, and so do not contribute to the temperature dependent zero-bias conductance.

The key findings of our investigations are (1) position-independent and (2) gapped tunneling and c-axis optical spectra at all temperatures, (3) metallic in-plane optical and dc conductivity, (4) c-axis dc conductivity which follows an activated form consistent with both STS and $\sigma_c(\omega)$ at high T but then tracks the ab-plane dc conductivity rather than the STS zero-bias conductance at low T, and (5) atomic resolution that is stable and reproducible but limited to very small regions on the extremely flat surfaces that we have prepared in ultra-high vacuum. All indications from the STM data themselves - including the step heights and lattice parameters measured when atomic resolution is found - demonstrate that the results are intrinsic to cleaved surfaces of LSMO. Furthermore, we have presented detailed evidence and analysis indicating that the STS, optical and dc conductivity measurements are entirely consistent with each other. This leaves us with the challenge of understanding the underlying anisotropy in quasiparticle transport out of the planes in LSMO and the seemingly astonishing observation (5).

The quasiparticles in strongly correlated materials can be very different and more complex than the renormalized electrons found in simple metals. Since STS extracts or inserts single (bare) electrons, the quasiparticles must dissociate in the tunneling process. Even if the quasiparticles form a metal in the plane, STS will measure a gap corresponding to the energy it takes to destroy the quasiparticle to extract the electron (hole) from the MnO layer. Thus, when the transport, STM, and optical data on LSMO are taken together, we are left with a picture where charge carriers are confined to metallic bilayers in LSMO, and cannot escape without a finite energy transfer. This is true down to the lowest temperatures, where our LSMO samples are magnetically ordered, and the conventional theory of solids would suggest a fully three-dimensional metal. The charge carriers are therefore bound to larger objects which can move within but not between the planes. Clear candidates for such objects are the polarons whose effects on the lattice and spin texture have been seen using X-rays and neutron scattering.[23,24] Polarons have also been associated with the pseudogap observed by ARPES[25] in LSMO with $x=0.4$ and the gap measured by STM[26] in the pseudo-cubic compound $La_{1-x}Ca_xMnO_3$ with $x=0$ and $x=0.3$.

It turns out that the same polaron picture provides a similarly clear understanding of observation (5) and is at the heart of our discovery of atomic-scale features in a metallic managanite. Specifically, the general inability to resolve atoms – in contrast to the case for the charge-ordered manganite $(Bi,Ca)MnO_3$ (Ref.27) where the valence fluctuations are slow or frozen – suggests metallic screening of charges within the ab-planes, in agreement with the bulk in-plane conductivity data. Indeed, for a standard free electron metal, screening reduces atomic contrast to below 0.02 Å for the tunneling parameters of our measurements.[22] Typically, STM can only image metal atoms (e.g. in gold) under the condition where there is bonding between the tip and sample and therefore actual relative motion of the atoms in question. Our experiments were performed outside this regime and are therefore insensitive. The small areas where atomic resolution is attained must therefore be regions where the effect of screening is relieved, due perhaps to a point defect. In some very famous experiments[28] on much simpler metals, defects result in quasiparticle scattering which is visible in the form of electron density modulations with periodicities fixed by the quasiparticle dispersion relation. Such effects have also been identified in the high temperature cuprates.[29] Obviously, if its potential is sufficiently strong, it will also be possible for a defect to bind (trap) a quasiparticle, resulting in a static (non-dispersive) STM pattern reflecting the internal structure of the polaron. There is also some experimental evidence for this phenomenon in the cuprates[30] at lower energies than those where quasiparticle dispersion effects dominate. We suggest here that this is what is occurring when we see atoms in LSMO, namely our experiments have captured the first real space image of that very elusive object – a trapped polaron. The contrast is derived from the charge and lattice distortions associated with a single polaron. The trapped polaron concept accounts not only for the relative paucity of regions with resolved atoms on surfaces which were optimized for smoothness and cleanliness, but also accounts for the remarkable standardization of their size, which actually coincides with neutron and X-ray scattering results[23,24] for the lattice and spin textures associated with polarons.

1. Anderson, P. W. The resonating valence bond state in $La_2CuO_4$ and superconductivity. *Science* **235**, 1196-1198 (1987).

2. Renner, Ch., Revaz, B., Genoud, J. Y., Kadowaki, K. & Fischer, Ø. Pseudogap precursor of the superconducting gap in under- and overdoped $Bi_2Sr_2CaCu_2O_{8+\delta}$. *Physical Review Letters* **80**, 149-152 (1998).

3. Uchida, S., Tamasaku, K. & Tajima, S. c-axis optical spectra and charge dynamics in $La_{2-x}Sr_xCuO_4$. *Physical Review B* **53**, 14558-14574 (1996).

4. von Helmolt, R., Wecker, J., Holzapfel, B., Schultz, L. & Samwer, K. Giant negative magnetoresistance in perovskitelike $La_{2/3}Ba_{1/3}MnO_x$ ferromagnetic-films. *Physical Review Letters* **71**, 2331-2333 (1993).

5. Jin, S. et al. Thousandfold change in resistivity in magnetoresistive La-Ca- Mn-O films. *Science* **264**, 413-415 (1994).

6. Millis, A. J. Lattice effects in magnetoresistive manganese perovskites. *Nature* **392**, 147-150 (1998).

7. Ramirez, A. P. et al. Thermodynamic and electron diffraction signatures of charge and spin ordering in $La_{1-x}Ca_xMnO_3$. *Physical Review Letters* **76**, 3188-3191 (1996).

8. Mori, S., Chen, C. H. & Cheong, S.-W. Pairing of charge-ordered stripes in $(La,Ca)MnO_3$. *Nature* **392**, 473-476 (1998).

9. Tokura, Y. & Nagaosa, N. Orbital physics in transition-metal oxides. *Science* **288**, 462-468 (2000).

10. Perring, T. G., Aeppli, G., Kimura, T., Tokura, Y. & Adams, M. A. Ordered stack of spin valves in a layered magnetoresistive perovskite. *Physical Review B* **58**, R14693-R14696 (1998).

11. Dagotto, E. *Nanoscale phase separation and colossal magnetoresistance* (eds. Cardona, M. et al.) (Springer, 2003).

12. Kimura, T. & Tokura, Y. Layered magnetic manganites. *Annual Reviews of Material Science* **30**, 451-474 (2000).

13. Abrikosov, A. A. Quantum interference effects in quasi-two-dimensional metals. *Physical Review B* **61**, 7770-7774 (2000).

14. Moritomo, Y., Asamitsu, A., Kuwahara, H. & Tokura, Y. Giant magnetoresistance of manganese oxides with a layered perovskite structure. *Nature* **380**, 141-144 (1996).

15. Becker, T. et al. Intrinsic inhomogeneities in manganite thin films investigated with scanning tunneling spectroscopy. *Physical Review Letters* **89**, 237203 (2002).

16. Fäth, M. et al. Spatially inhomogeneous metal-insulator transition in doped manganites. *Science* **285**, 1540-1542 (1999).

17. Kimura, T. et al. Interplane tunneling magnetoresistance in a layered manganite crystal. *Science* **274**, 1698-1701 (1996).

18. Kittel, C. *Introduction to solid state physics* (John Wiley and Sons, Inc., New York, 1976).

19. Ishikawa, T., Tobe, K., Kimura, T., Katsufuji, T. & Tokura, Y. Optical study on the doping and temperature dependence of the anisotropic electronic structure in bilayered manganites: $La_{2-2x}Sr_{1+2x}Mn_2O_7$ ($0.3 <= x <= 0.5$). *Physical Review B* **62**, 12354-12362 (2000).

20. Freeland, J. W. et al. Full bulk spin polarization and intrinsic tunnel barriers at the surface of layered manganites. *Nature Materials* **4**, 62-67 (2005).

21. Konoto, M. et al. Direct imaging of temperature-dependent layered antiferromagnetism of a magnetic oxide. *Physical Review Letters* **93**, 107201 (2004).

22. Blanco, J. M. et al. First-principles simulations of STM images: From tunneling to the contact regime. *Physical Review B* **70**, 085405 (2004).

23. Perring, T. G., Aeppli, G., Moritomo, Y. & Tokura, Y. Antiferromagnetic short-range order in a two dimensional manganite exhibiting giant magnetoresistance. *Physical Review Letters* **78**, 3197-3200 (1997).

24. Vasiliu-Doloc, L. et al. Charge melting and polaron collapse in $La_{1.2}Sr_{1.8}Mn_2O_7$. *Physical Review Letters* **83**, 4393-4396 (1999).

25. Dessau, D. S. et al. *k*-dependent electronic structure, a large "ghost" Fermi surface, and a pseudogap in a layered magnetoresistive oxide. *Physical Review Letters* **81**, 192-195 (1998).

26. Wei, J. Y. T., Yeh, N. C. & Vasquez, R. P. Tunneling evidence of half-metallic ferromagnetism in La0.7Ca0.3MnO3. *Physical Review Letters* **79**, 5150-5153 (1997).

27. Renner, Ch., Aeppli, G., Kim, B. G., Soh, Y. A. & Cheong, S.-W. Atomic-scale images of charge ordering in a mixed-valence manganite. *Nature* **416**, 518-521 (2002).

28. Crommie, M. F., Lutz, C. P. & Eigler, D. M. Imaging standing waves in a 2-dimensional electron-gas. *Nature* **363**, 524-527 (1993).

29. Hoffman, J. E. et al. Imaging quasiparticle interference in $Bi_2Sr_2CaCu_2O_{8+\delta}$. *Science* **297**, 1148-1151 (2002).

30. Yazdani, A., Howald, C. M., Lutz, C. P., Kapitulnik, A. & Eigler, D. M. Impurity-induced bound excitations on the surface of $Bi_2Sr_2CaCu_2O_8$. *Physical Review Letters* **83**, 176-179 (1999).

We are grateful to T.Ishikawa for providing the optical conductivity data shown in Figure 4. We thank D.McPhail and R.Chater for the static SIMS measurements and A.Fisher and J.Mesot for very stimulating discussions. We acknowledge support from a Wolfson-Royal Society Research Merit Award, the NEC corporation and the European Commission through a FW6 STREP programme, and one of us (HMR) thanks T.F.Rosenbaum of the University of Chicago for support and encouragement.

Correspondence and requests for materials should be addressed to C. R. (e-mail: c.renner@ucl.ac.uk).

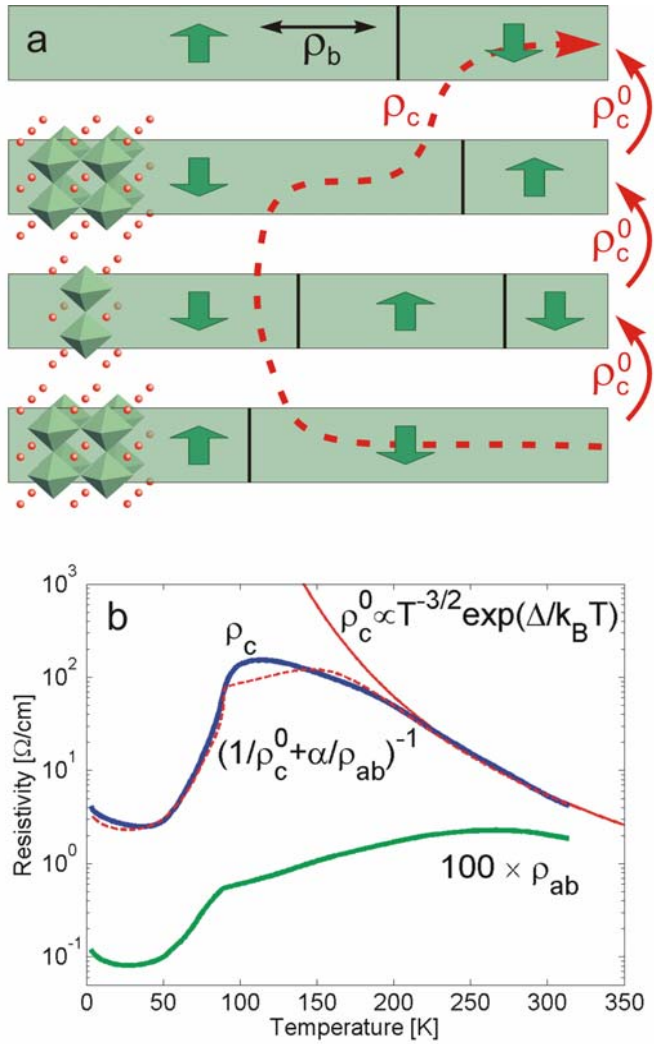

Figure 1 Structure and bulk transport properties of $La_{1.4}Sr_{1.6}Mn_2O_7$. (a) Bilayer perovskite structure and different paths considered for the c-axis transport. The green arrows depict the ferromagnetic domains, and the dashed red line a possible reduced resistance c-axis conduction path (see text for details). (b) The measured in-plane resistivity, $\rho_{ab}$, (green) and c-axis resistivity, $\rho_c$, (blue) drop dramatically below the ferromagnetic ordering temperature. The high temperature part of the c-axis resistivity fits a purely activated behaviour (red), $\rho_c^0(T) \propto T^{-3/2} exp(\Delta_\rho/2k_BT)$, with $\Delta_\rho = 188$ meV consistent with the full gap $\Delta = V_+ - V_-$ found by STM in this temperature range. The dashed red line depicts the resistivity obtained from the model described in the text, including the in-plane contributions and conductivity enhancements at magnetic stacking faults as illustrated in **a**.

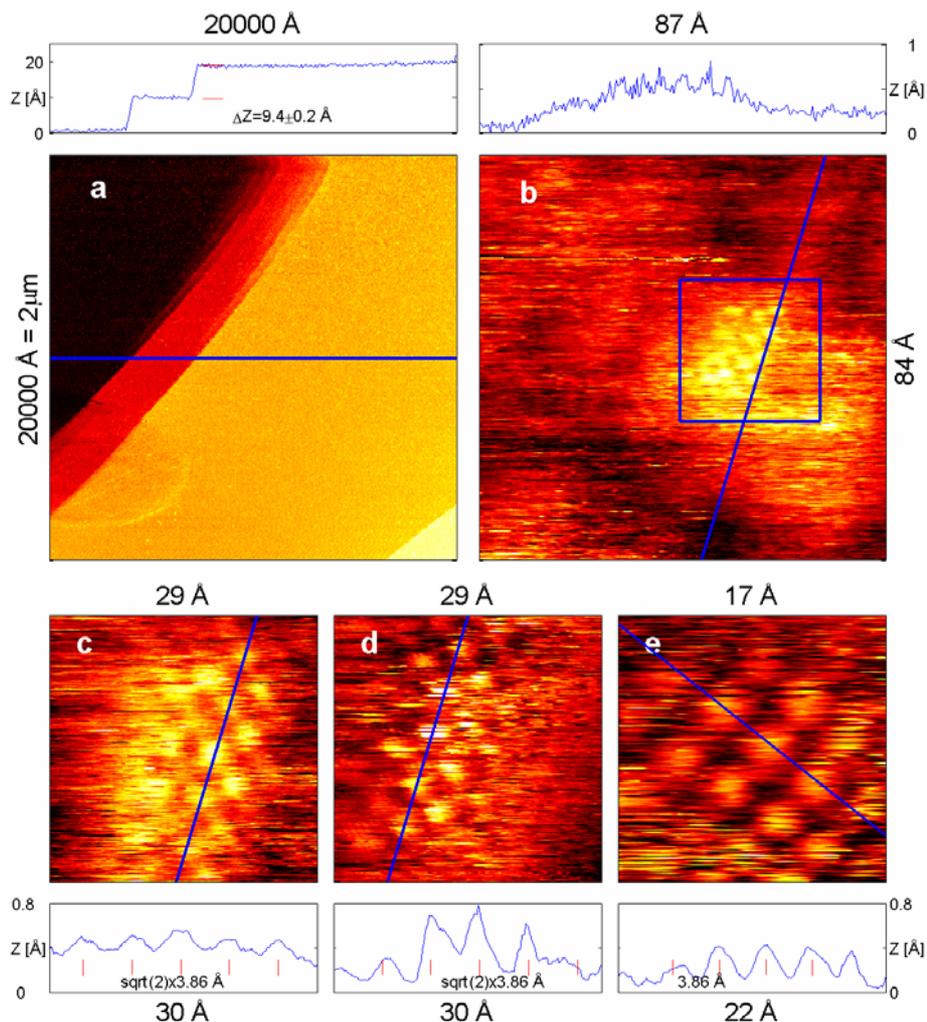

Figure 2 Raw STM micrographs of $La_{1.4}Sr_{1.6}Mn_2O_7$. ($V$=0.8V, $I$=0.2nA). The sample is held at a positive potential relative to the tip, implying that we are primarily probing the unoccupied electronic density of states of the manganite. (a) In-situ cleaving exposed atomically flat micron-sized terraces (imaged at 212 Kelvin). On this scale, the non-linear response of the piezoelectric STM scanner accounts for the curved appearance of the steps. (b,c) Higher magnification images of the same cluster obtained in separate runs. Atomic scale resolution is always restricted to tiny regions, containing about 30 unit cells (imaged at 283 Kelvin). (d,e) Two other instances of tiny regions with atomic resolution. These regions are systematically associated with a locally enhanced conductance at 0.8V. The amplitude profiles show selected cross-sections from each micrograph (blue lines) highlighting the half unit-cell high steps **a** and the enhanced tunneling conductance **b** associated with atomic-scale resolution **c-e**.

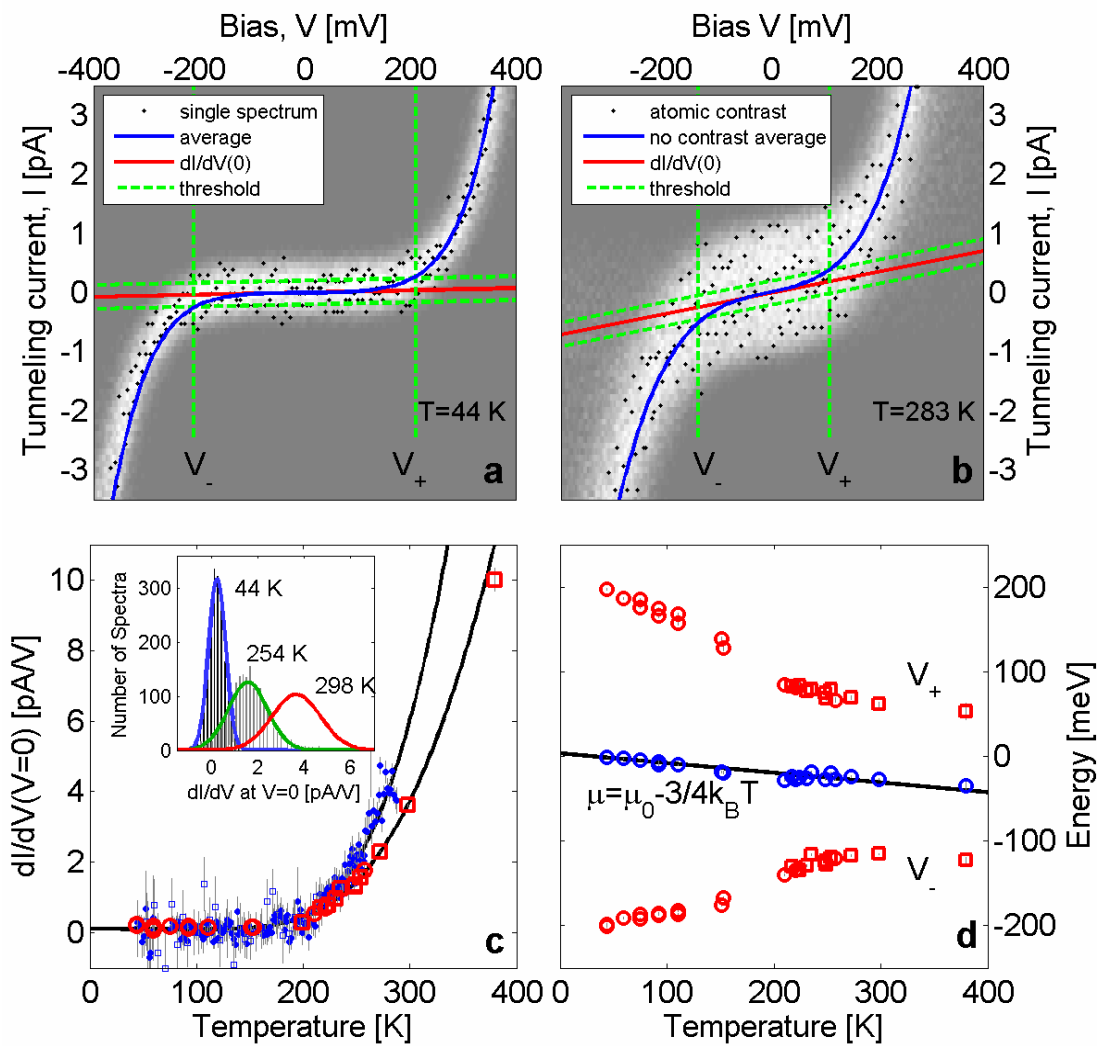

Figure 3 Temperature-dependent STM spectroscopy of $La_{1.4}Sr_{1.6}Mn_2O_7$. (a) Current-voltage characteristics at 44 Kelvin, and (b) at 283 Kelvin (see text for further details). (c) Tunneling conductance $\sigma_0$ at $V=0$ as a function of temperature. The open symbols are obtained from the averages of thousands of $I(V)$ spectra measured over large areas at fixed temperatures, while the solid symbols derive from single $I(V)$ characteristics measured on slowly warming. Fitting $\sigma_0(T)$ to $T^{3/2} exp(-\Delta/2k_BT)$ gives $\Delta=196\pm12$meV for both sets of data. The inset shows the distribution of $\sigma_0$ obtained from thousands of $I(V)$ characteristics measured at individual points on large areas at selected fixed temperatures. The uniform distributions exclude any metal-insulator phase separation at all temperatures. (b) Temperature dependence of the gap edges $V_-$ and $V_+$, showing the persistence of the gap below the metal-insulator transition at 90 Kelvin. The blue circles feature the mid-gap energy ½($V_+$ + $V_-$ ) which is independent of the chosen threshold (0.2pA in this case) and corresponds to the chemical potential $\mu$.

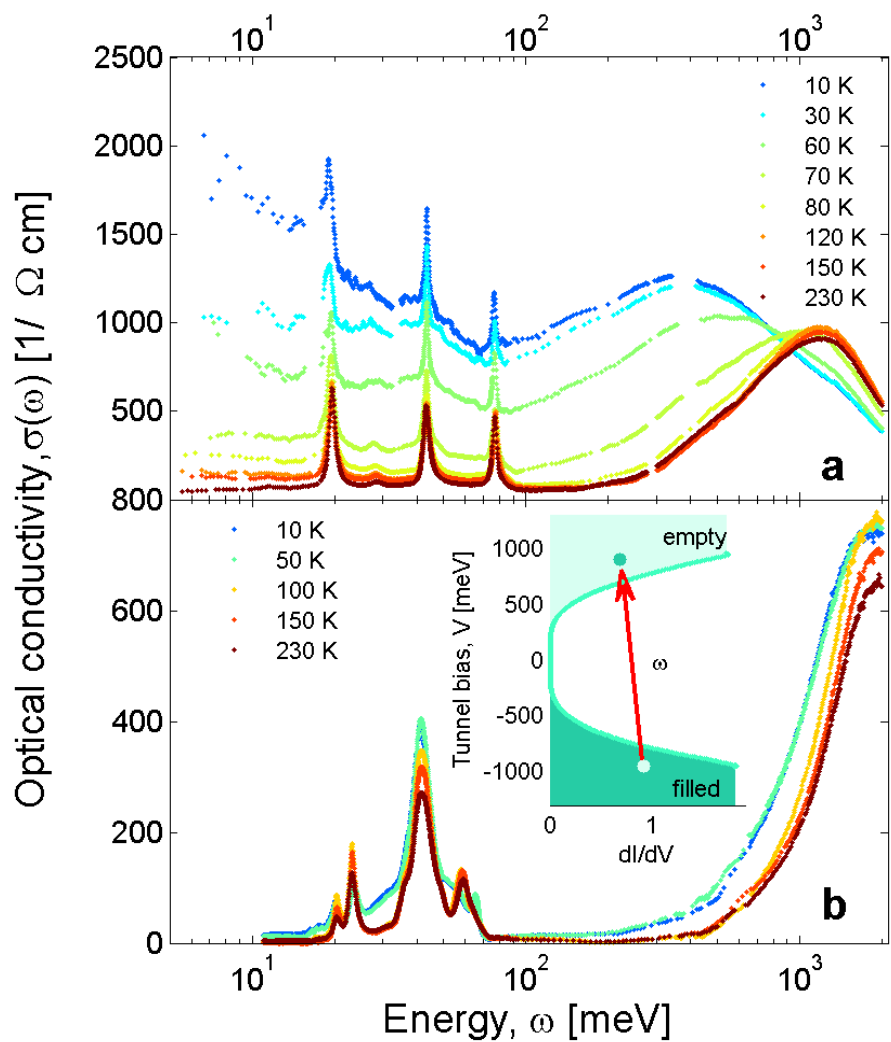

Figure 4 In-plane and out-of-plane optical conductivity deduced from near-normal-incidence reflectivity measured on single crystal samples. Detailed experimental and analytical procedures for the optical measurements are described elsewhere.[19] (a) In-plane optical conductivity. Sharp peaks at 20-80 meV are due to phonons. The dramatic increase at low frequencies below 90 Kelvin corresponds to Drude-like metallic behaviour. (b) The c-axis optical conductivity, which except for the 20-70 meV phonon bands, only shows a semiconductor-like absorption edge at around 400 meV. Inset shows how the conductivity is related to the single particle density of states measured via STM (Figs. 3a and 3b). Imperfect alignment of the photon polarization and the crystal c-axis mixes the measured in-plane and out-of-plane conductivities, and is the most likely cause of the reduction of the absorption edge on cooling.